# Option Pricing Using Bayesian Neural Networks

*Michael Maio Pires, Tshilidzi Marwala*

School of Electrical and Information Engineering, University of the Witwatersrand, 2050, South Africa

m.pires@ee.wits.ac.za, t.marwala@ee.wits.ac.za

## Abstract

Options have provided a field of much study because of the complexity involved in pricing them. The Black-Scholes equations were developed to price options but they are only valid for European styled options. There is added complexity when trying to price American styled options and this is why the use of neural networks has been proposed. Neural Networks are able to predict outcomes based on past data. The inputs to the networks here are stock volatility, strike price and time to maturity with the output of the network being the call option price. There are two techniques for Bayesian neural networks used. One is Automatic Relevance Determination (for Gaussian Approximation) and one is a Hybrid Monte Carlo method, both used with Multi-Layer Perceptrons.

## 1. Introduction

This document deals with the use of two kinds of Bayesian neural networks applied to the American options pricing problem. Both Bayesian techniques used were used with Mult-Layer Perceptron (MLP) networks. The techniques can also be used with Radial Basis Function (RBF) networks [1] but they were only used with MLP networks here. The two Bayesian techniques used are Automatic Relevance Determination (ARD) (for Gaussian Approximation) and the Hybrid Monte Carlo method (HMC) which will be discussed.

Firstly we need to introduce the notion of an option. An option is the right (not the obligation) to buy or sell some underlying asset at a later date but by fixing the price of the asset now [2]. For someone to have this option, he/she has to pay a fee known as the option price. There are two kinds of options, namely a call and a put option. A call option gives the person the right to buy the underlying asset and a put option gives the person the right to sell the underlying asset [2]. The pricing of either call or put options is equally difficult and something that has brought much research interest.

Black et al. [3] provided equations in 1973 that provided a pricing formula for call and put options. To obtain these equations, several assumptions had to be made. The most important assumption made is that the formulas only held for European styled options [4]. European styled options only allow the exercise of the option on the maturity date (which is the later date that the person is allowed to buy or sell the underlying asset) [5]. What are used extensively worldwide, though, are American styled options where the person is allowed to buy or sell the underlying asset at any date leading up to the maturity date. This introduces another random process into the pricing of the option (because it cannot be predicted when the exercise of the option will occur) and so the pricing of these kind of options is much more complex than European styled options [6].

Neural Networks (NN's) are a form of prediction based on trends that have occurred in the past. The outputs of the network are that which are to be predicted and the inputs are chosen as variables that affect the outputs in the real world and whose trends can be used to predict the output variables. MLP and Support Vector Machines (SVM's) have been used to price American options [7] and here what will be tested is the effectiveness of Bayesian Neural Networks.

## 2. Bayesian Neural Networks

### 2.1 Bayesian Techniques

With NN's there is always an error in the predictions made and we thus have

$$y = f(x; w) + \varepsilon \qquad (1)$$

where $y$ is the actual output desired, $f$ is the output predicted by the network, $\varepsilon$ is the error, $w$ are the weights [1] and $x$ is a vector of inputs. Even if we are given $\varepsilon$ and the same network is run twice with the same parameters, we will obtain different weights $w$ both times and thus there is an uncertainty in the training of the networks [1] and this can be attributed to the randomness in the assignment of weights. Generally some complex models try to fit the noise into the predictions which cause problems when trying to predict with unseen inputs (the problem of over training) and thus cause there to be even more error in the predictions [1].

*p(.)* wherever used from now on is used to denote the probability function from statistics. In the Bayesian approach, the uncertainty in the parameters estimated when training a network is assumed to follow a particular distribution. We first start with a *prior* distribution *p(w)* which gives us an idea of the parameters before the data is used [1] but this only give us a vague idea as the distribution is quite broad. The prior distribution can be of any kind for example Poisson or Geometric. In this case we will only use a Geometric distribution. We then wish to narrow this distribution down by finding the posterior probability density of the parameters *w* given a particular dataset *D*, *p(w|D)* where

$$p(w|D) = \frac{p(D|w)p(w)}{p(D)} \quad (2)$$

and *p(D|w)* is the dataset likelihood and *p(D)* is the evidence and ensures that the posterior integrates to 1 and is calculated by an integral over the parameter space. Once the posterior

$$p(D) = \int p(D|w')p(w')dw' \quad (3)$$

is calculated we can then make a prediction at a new input by first calculating the prediction distribution

$$p(y|x^*, D) = \int p(y|x^*, w)p(w|D)dw \quad (4)$$

where *y* is the predicted values and then the actual prediction is found by

$$E(y|x^*, D) = \int y p(y|x^*, w)p(w|D)dw \quad (5).$$

*E(.)* is the expected value in statistical terms. As can be seen from equations (3) and (5), there is an integral involved and the dimensionality of the integral is given by the number of network parameters (weights) and this is not analytically possible and simple numerical algorithms break down [1]. Therefore approximations to the posterior are made (the toolbox used to train Bayesian Neural Networks is the NETLAB toolbox used with MATLAB®) and this is known as the evidence function in NETLAB and is used together with a Gaussian Approximation and ARD (see section IIB). What can also be used is Hybrid Monte Carlo (HMC) methods combined with Monte Carlo sampling used for integral approximation [1] (see section IIC).

The main reason for the use of Bayesian techniques is simply to reduce the uncertainty in the weights and thus try to reduce the problem of over fitting (i.e. over fitting occurs when a network predicts badly because it is trained too much to its training data and predicts badly with unseen inputs [1]). Bayesian techniques do reduce the problem of over fitting as has been proved by Nabney [1]. In NN's there is a need to optimize the network and thus reduce the error function [8]. In Bayesian techniques this is done by obtaining a posterior distribution for the weights so that they can only be found within a particular distribution thus narrowing the search for the optimal weight values [1]. Bayes' theorem helps us do this but there are large integrals and there are several ways of evaluating these integrals. There are Gaussian Approximations and HMC.

## 2.2 Automatic Relevance Determination

The prior distribution is chosen to be Gaussian [1] and thus is of the form

$$p(w) = \frac{1}{Z_W(\alpha)} e^{-\frac{\alpha}{2}\sum_{i=1}^{W} w_i^2} \quad (6)$$

where the normalization constant $Z_W(\alpha)$ is

$$Z_W(\alpha) = \left(\frac{2\pi}{\alpha}\right)^{W/2} \quad (7).$$

*α* is known as the hyperparameter because it is a parameter for the distribution of other parameters. It is then helpful to have different hyperparamaters, one for each set of the weight sets $W_1..., W_g$. The way to choose these different hyperparamters is to have values for them associated to how important each input variable is. This is known as Automatic Relevance Determination (ARD).

ARD is used because there is often the need to find the relevance of certain input variables. This is not easily done if there are hundreds of input variables. In Bayesian NN's we associate each hyperparameter with an input variable. Each hyperparameter represents the inverse variance of the weights and so the lower the value for a hyperparameter associated with a particular input, the more important that input is in the prediction process because it means that large weights are allowed [1].

## 2.3 Hybrid Monte Carlo Method

As stated before, Monte Carlo methods can be used to approximate the integrals involved in Bayesian techniques rather than using a Gaussian approximation with ARD and an evidence procedure [1].

Since there is an uncertainty in the process, we need to find the predictive distribution, i.e. the distribution that represents the possible outcomes of the network due to the uncertainty in the weights [1]. This distribution is an integral but in Monte Carlo methods it is approximated to a sum

$$p(y|x,D) = \frac{1}{N}\sum_{n=1}^{N} p(y|x,w_n) \qquad (8).$$

where *N* is the number of samples chosen by the trainer of the network and $w_n$ is the sample of weight vectors. These samples of weights can be chosen through different methods. A Metropolis-Hastings algorithm can be used to sample these weights but has proved to be very slow this is because the method makes no use of gradient information and for NN's the method of error back-propagation provides an algorithm for evaluating the derivative of an error function and thus optimizing the network more computationally efficiently [1]. Another method that can be used is the Hybrid Monte Carlo (HMC) algorithm for sampling which is the one that is used in this application and makes use of the gradient information.

The HMC algorithm is a sampling algorithm that takes into consideration certain gradient information. The algorithm follows the following sequence of steps once a step size *ε* and the number of iterations *L* has been decided upon:

*1) Randomly Choose a Direction λ:* λ can be either -1 or +1 with the probability of either being chosen being equal.

*2) Carry Out the Iterations:* Starting with the current state $(w, p) = (\hat{w}(0), \hat{p}(0))$ randomly selected, where *p* is a momentum term which is evaluated at each step, we then perform *L* steps with a step size of *λε* resulting in the candidate state $(\hat{w}(\lambda\varepsilon L), \hat{p}(\lambda\varepsilon L)) = (w^*, p^*)$.

*3)* The candidate state is accepted with probability $\min(1, e^{(-H(w^*,p^*)-H(w,p))})$ where *H(.)* is the Hessian matrix. If the candidate state is rejected then the new state will be the old state.

These three steps, in essence, describe how the sampling is done so that the summation of equation (8) can be accomplished and so that the posterior distribution can be found and thus allowing the optimization of the NN. The momentum term *p* can be randomly generated or it can be changed dynamically at each step and there are different ways of doing this [9]. The sets of weights are thus selected or rejected according to the three steps above and the number of samples that are wished to be retained are the number of weights retained. For each set of weights there is a corresponding NN output. The prediction of the network is the average of the outputs.

The usefulness of the Bayesian approach comes into the fact that the prediction comes with certain confidence levels. In fact the prediction mathematically is the same as that of the standard MLP. If we plot the prediction and upper and lower bounds (where the upper bound is the prediction plus the standard deviation of the outputs and the lower bound is the prediction minus the standard deviation of the outputs of the network) then we say that the prediction is known to within a certainty of 68% (because in the normal distribution 1 standard deviation form the mean constitutes 68% of the possible outcomes [10]). This is done for the Gaussian and HMC approaches.

## 3. Results of Bayesian Neural Networks

### 3.1 Automatic Relevance Determination Approach

Data was obtained from the JSE Securities Exchange of South Africa. It was obtained for a particular stock option for the period January 2001 to December 2003. This resulted in there being 3051 points of data that could be used for training and testing of the networks trained. The inputs to the network were stock volatility, strike price and time to maturity (in days). The output of the network would simply be the call price of an option. Call prices were obtained for different options with there being both high and low prices. What was decided was to use the average of the high and low prices as the actual call price and these are the values used to train and test the network.

There are demos available in the NETLAB toolbox that show the procedure of training Bayesian NN's with the Gaussian Approximation and ARD, and HMC. These demos were edited so that the procedures could be experimented with on the options pricing problem. In the Gaussian Approximation with ARD, it was found that 500 training cycles showed the best results with 1000 data points being used to train the network. The networ k was tested with 300 data points so that the plots could be easily seen when viewing the error bars. The evidence procedure utilized in the toolbox has a certain amount of cycles associated with it as well and it was found that 10 cycles for this sufficed for the training of the Bayesian NN. The parameters changed were the number of hidden units, the number of loops used to find better hyperparameter values and the value for *β* that is associated with MLP NN's and is the coefficient of data error associated with the MLP. The results of the Gaussian Approximation approach with ARD can be seen in table I.

There was a problem when trying to find the standard deviations of the outputs for the Bayesian NN's using the ARD approach. The function that provides the standard deviations, at times, produced some imaginary numbers so what was done was to search through the standard deviations and replace the imaginary numbers with the first standard deviation value in the array. This got rid of the errors in MATLAB® but showed that the ARD approach does have some bugs. In fact it is said that the Gaussian approximation is the same as the HMC under certain conditions but these conditions are not known and in fact the only reason that Gaussian approximations are used in Bayesian techniques is because they are more mathematically neat than other Bayesian approaches.

As can be seen from table I, the network performed the best with the coefficient of data error at 10, with 50 hidden units and the number of loops to find different hyperparameter values only set to 1. The values found for the different hyperparameters show that each input was important in the determination of call prices because each hyperparameter was in the same order of magnitude and there isn't one that is significantly smaller or

## TABLE I
### ARD RESULTS

| β | Hid. Units | Mean Error (%) | Time (s) | n | σ | Alphas |
|---|---|---|---|---|---|---|
| 1 | 25 | 64.7 | 52 | 1 | 1516 | [1.2177 1.2036 0.6417] |
| 10 | 25 | 53.16 | 52 | 1 | 1512 | [0.8366 0.9051 0.5723] |
| 100 | 25 | 61.94 | 49 | 1 | 1354 | [1.0101 0.9760 0.4525] |
| 1 | 50 | 57.44 | 104 | 1 | 1541 | [1.2231 0.9342 1.1528] |
| 10 | 50 | 52.72 | 103 | 1 | 1505 | [1.5385 0.8931 0.8203] |
| 100 | 50 | 61.16 | 102 | 1 | 1485 | [0.7763 1.2248 0.9373] |
| 1 | 25 | 62.1 | 105 | 2 | 1390 | [1.7646 0.9891 1.1858] |
| 10 | 25 | 70.49 | 97 | 2 | 1520 | [1.8534 0.7471 0.9004] |
| 100 | 25 | 58.95 | 102 | 2 | 1433 | [1.1214 1.5703 0.4662] |
| 1 | 50 | 78.49 | 191 | 2 | 1521 | [2.0210 1.5175 0.7859] |
| 10 | 50 | 76.56 | 177 | 2 | 1409 | [1.7518 1.2180 0.6775] |
| 100 | 50 | 61.85 | 175 | 2 | 1456 | [1.9264 1.3430 0.7552] |

β = coefficient of data error for the MLP, Hid. Units = number of hidden units used in the training of the MLP, Mean Error = average error found by subtracting each prediction from the actual value and multiplying by 100 over the size of the test set used (300), Time = time taken to train the network, n = number of loops used to find the best hyperparameter values, σ = the average size of the bounds for all the outputs (average of standard deviations of output samples), Alphas = hyperparameter values found for the corresponding input to hidden unit weights thus showing the importance of the different inputs.

larger than the others. The time column indicates that the networks didn't take too long to train and that if the number of hidden units was doubled so the time to produce a result also doubled (give or take a few seconds). Other values were tried for hidden units and also what was also tried was to use more training data to improve the accuracy of the pricing model. It was found that with 1500 training points and 100 hidden units the mean error was much higher than the values found in table I and also took up to 30 minutes to train. Note that to obtain these results the algorithm had to be run several times with the same parameters so that the best results for these parameters could be obtained, this is due to the random nature inherent in the algorithm for training the NN as was found with standard MLP's [7]. The standard deviations found for each network trained are quite large and thus the predictions found by the network are known to be within a range of about R3000 with a certainty of only 68%. Therefore we can only say that we know the price to be within quite a large range (of R3000) and only with a confidence of 68%. The outputs for 100 of the 300 test points used and with the corresponding confidence levels for the 2$^{nd}$ network in table I can be seen in Figure 1.

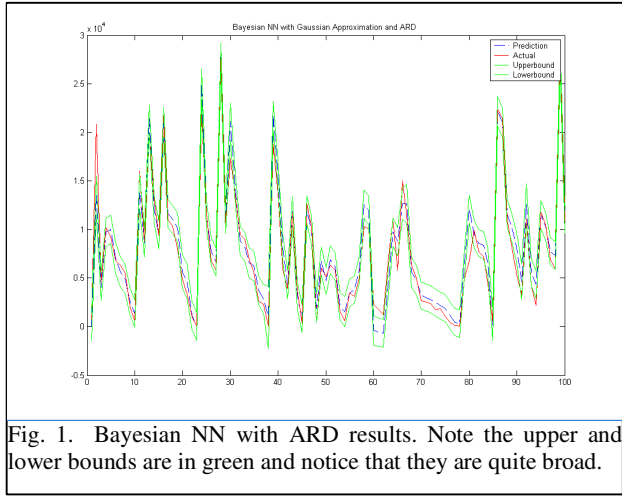

Fig. 1. Bayesian NN with ARD results. Note the upper and lower bounds are in green and notice that they are quite broad.

### 3.2 Monte Carlo Approach

The data used to train and test the HMC Bayesian NN was the same data as that used for the ARD approach. Here the coefficient of data error value was not experimented with and was rather kept at a value of 10. The number of hidden units was experimented with as well as the number of initial samples rejected and the number of samples in the HMC procedure. The step size was kept constant at 0.002 because it was found that if it was changed to other values bigger or smaller then the threshold (probability used in the rejection criteria) was not a number (NaN in MATLAB®) and so the procedure didn't work very well in these cases. The number of training points used was also 1000 and the number of points used to test the network was 300. The results for the HMC Bayesian NN approach can be seen in table II.

As can be seen from table II, the networks took quite sometime to train with 1000 training points. It was attempted to try fewer points for training but just reduced the performance of the network significantly. What was also attempted was to use more hidden units to train the network but this proved to increase the amount of time required to train the network with no improvement in the error analysis. Note that the algorithm for each result in table II was found by training the same network only once. It didn't have to be run several times. The process of training networks in this was is still random but the seed used for the random number generator was the same every time and so there was no difference between the results of two networks that were trained with the same parameters. The standard deviations found by each network trained are significantly smaller than that found by the Gaussian approach

| TABLE II HMC RESULTS | | | | | | |
|---|---|---|---|---|---|---|
| Rej. | Max Error (%) | Mean Error (%) | Samp. | Time (s) | σ | Hidden Units |
| 100 | 5241 | 76.07 | 100 | 259 | 445.95 | 10 |
| 100 | 5990 | 95.68 | 100 | 444 | 502.72 | 20 |
| 100 | 4372 | 82.76 | 100 | 816 | 699.36 | 40 |
| 100 | 4378 | 98.31 | 400 | 648 | 468.71 | 10 |
| 100 | 5212 | 77.92 | 400 | 1114 | 575.67 | 20 |
| 100 | 6719 | 98.26 | 400 | 2104 | 814.61 | 40 |
| 200 | 5662 | 79.21 | 100 | 390 | 401.42 | 10 |
| 200 | 7618 | 103.42 | 100 | 665 | 684.75 | 20 |
| 200 | 4021 | 91.80 | 100 | 1227 | 680.83 | 40 |
| 200 | 3849 | 92.04 | 400 | 777 | 472.08 | 10 |
| 200 | 4093 | 78.29 | 400 | 1322 | 591.20 | 20 |
| 200 | 5836 | 78.53 | 400 | 2451 | 722.30 | 40 |

Rej. = number of samples to be rejected initially (at the start of the Markov chain), Max Error = Maximum error between the actual output and that predicted by the network in the 300 point test set used, Mean Error = average error of the size of the test set used (300), Samp. = number of samples in the HMC method, Time = time taken to train the network, σ = the average size of the bounds for all the outputs (average of standard deviations of output samples), Hidden Units = number of hidden units used in the MLP.

with ARD. Therefore the predictions of the network are known with a confidence of also 68% to be within a certain range but the range is much smaller and at best the range was R802.84. The outputs for 100 of the 300 test points used and with the corresponding confidence levels for the 1st Network in table II can be seen in Figure 2.

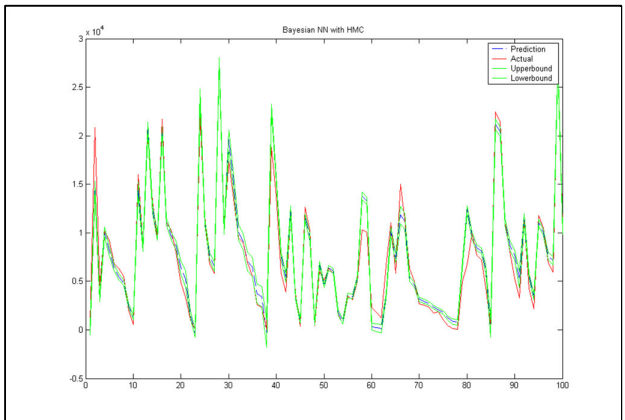

Fig. 2. Bayesian NN with HMC results. Note the upper and lower bounds are in green and notice that they are much less broad than the bounds for the ARD approach.

## 4. Comparison of Bayesian Techniques with Standard Multi-Layer Perceptrons and Support Vector Machines

From the results obtained for the standard MLP and SVM [7], it must be said that the Bayesian techniques applied to NN's didn't provide any improvements. In fact mathematically they are said to be the same as standard NN's but the advantage they bring is the actual confidence levels. With regards to the ARD approach, the best level of mean error was found to be 53% which is very close to the 51% found by the standard MLP trained before. The amount of time taken to train the network was much more than that found by the standard MLP as was to be expected due to the extra functions being utilized in the Bayesian approach due to the approximations inherent in the technique. Compared to SVM it was faster than the 7 minutes taken to train an SVM network but the results were significantly poorer because the average error found by the SVM network at best was 34.4%.

With regards to the HMC approach the best value found for average error over the test set was found to be 76.07%. HMC is mathematically supposed to provide the same results as standard MLP's but it didn't in this case. This is probably because not enough samples were taken when obtaining a prediction. With there being 400 samples the network took up to 40 minutes to train and so for the purposes of this study what was considered to be more interesting is the fact that HMC provided a much narrower band of confidence than that found by the Gaussian approach with ARD. The band produced by the HMC approach was R804.84 which is significantly better than the R3000 found by the ARD approach. Therefore even though the error found by the HMC approach was found to have at best an average of 76.07% we know that the price given by the network is known to be within a band of R804.84 with a confidence of 68%. A drawback is of course the time taken to train the network using HMC. It takes very long but is still more useful than standard MLP's and MLP's with the ARD approach.

In conclusion the best NN method was found to be the SVM method because it produced the best error analysis results and even though it took 7 minutes to train it is worth using in the future. But it must be said that Bayesian NN's do produce confidence levels for the outputs which is still a serious advantage over standard NN's when pricing options. This is because what can be done is to say that a price is provided with this degree of confidence and thus we can then see the implications of adding a bit to the price because we know the confidence or subtracting from the price. Based on this we can se that optimally a Bayesian SVM approach would be favorable and this could be further researched.

## 5. Conclusion

The algorithm that worked the best for the option pricing problem is the SVM algorithm. It produced the best error analysis results even though it takes a bit longer to train than standard MLP NN's and Bayesian MLP NN's with ARD. What can be attempted in the future is to use some optimization approach (such as Particle Swarm Optimization or Genetic Algorithm) to obtain the optimum number of weights and values for other parameters so that the best Bayesian NN can be found. This may prove to be very computationally intensive and may take a very long time especially with the HMC approach with Bayesian NN's. Bayesian techniques can be very powerful and should be experimented with further so that the best parameters for them can be found but at first hand it has been found that the best performing NN is the SVM. The HMC Bayesian approach provides the best confidence levels and maybe a combination of these confidence levels with SVM can be attempted in some manner.